\documentclass[letterpaper,twocolumn,10pt]{article}
\usepackage{aicomps-2024-07}

\usepackage{tikz}
\usepackage{amsmath}

\usepackage{booktabs}
\usepackage{multirow}
\usepackage{graphicx}
\usepackage{threeparttable}
\usepackage{array}

\usepackage{pifont}

\begin{document}
%-------------------------------------------------------------------------------

%don't want date printed
\date{}

\title{\Large \bf A Review on Proprietary Accelerators for Large Language Models}

%for single author (just remove % characters)
\author{
{\rm Sihyeong Park}\\
Korea Electronics Technology Institute
\and
{\rm Jemin Lee}\\
Electronics and Telecommunications Research Institute
\and
{\rm Byung-Soo Kim}\\
Korea Electronics Technology Institute
\and
{\rm Seokhun Jeon}\\
Korea Electronics Technology Institute
% \author{
% {\rm Author 1}\\
% Affiliation 1
% \and
% {\rm Author 2}\\
% Affiliation 2
% \and
% {\rm Author 3}\\
% Affiliation 3
% \and
% {\rm Author 4}\\
% Affiliation 4
%Name Institution
} % end author

\maketitle

%-------------------------------------------------------------------------------
\begin{abstract}
%-------------------------------------------------------------------------------
With the advancement of Large Language Models (LLMs), the importance of accelerators that efficiently process LLM computations has been increasing. This paper discusses the necessity of LLM accelerators and provides a comprehensive analysis of the hardware and software characteristics of the main commercial LLM accelerators. Based on this analysis, we propose considerations for the development of next-generation LLM accelerators and suggest future research directions.

\end{abstract}

%-------------------------------------------------------------------------------
\section{Introduction}
%-------------------------------------------------------------------------------

Following the success of ChatGPT~\cite{achiam2023gpt}, research and development of Large Language Models (LLMs) and their applications have accelerated. Recently, multimodal LLMs~\cite{wu2023next, li2024llava} capable of handling text, images, audio, and video, beyond traditional text-based chatbots, have emerged. As these models grow in complexity, data requirements and associated hardware costs for training and inference have increased exponentially.

Graphics Processing Units (GPUs) are commonly used for LLM training and inference due to their versatility and availability. However, GPUs' high power consumption causes heat issues, necessitating costly cooling systems. Thus, optimizing hardware for LLM computations is essential to reduce costs.

LLM accelerators, designed specifically to enhance LLM computations, are being developed in various forms. Unlike general-purpose GPUs, LLM accelerators use dedicated circuits for efficiency, offering higher performance with lower energy consumption. While hardware advancements are crucial, supporting software stacks are also increasingly important.

In this paper, we analyze the hardware and software of LLM accelerators, particularly focusing on off-the-shelf LLM accelerators. Based on this, we discuss the essential elements and challenges for LLM accelerators. Specifically, we target commercial LLM accelerators implemented as Application-Specific Integrated Circuits (ASICs). We concentrate on accelerators that are accessible through research papers or white papers and are used in actual data centers. Our paper aims to analyze the characteristics of proprietary LLM accelerators, identify the challenges, and suggest future research directions.

% The structure of this paper is as follows: Section~\ref{sec:2} explains LLM accelerators, Section~\ref{sec:3} reviews proprietary LLM accelerators, Section~\ref{sec:4} discusses the challenges faced by LLM accelerators based on the preceding content, and conclusions are presented in Section~\ref{sec:5}.

%-------------------------------------------------------------------------------
\section{LLM Accelerators}
\label{sec:2}
%-------------------------------------------------------------------------------

% LLM accelerators are specialized hardware designed to enhance the performance and efficiency of LLMs. As the complexity and functionality of LLMs grow, the need for effective acceleration becomes more critical, especially with increasing demand in fields like natural language processing, speech recognition, and automation. LLM accelerators optimize processing speed and energy efficiency, thereby reducing latency and providing a better user experience.

% \footnote{Remember that AICompS format follows the USENIX format stopped using endnotes and is
%   now using regular footnotes.} And some embedded literal code may

%-------------------------------------------------------------------------------
\subsection{Types of Accelerators}
%-------------------------------------------------------------------------------

% LLM accelerators offer high efficiency based on hardware architectures optimized for LLM computations. Major types of LLM accelerators include GPUs, Tensor Processing Units (TPUs), ASICs, Field-Programmable Gate Arrays (FPGAs), and Neural Processing Units (NPUs).

LLM accelerators offer high efficiency based on hardware architectures optimized for LLM computations. Major types of LLM accelerators include GPUs, ASICs, Field-Programmable Gate Arrays (FPGAs).

\textbf{GPU} GPUs are the most widely used LLM accelerators, capable of processing vast amounts of data quickly due to their highly parallel structure. They can execute multiple threads simultaneously and enable parallel processing through multi-GPU interconnection technologies.

% \textbf{TPU, NPU} TPUs and NPUs are hardware designed to accelerate machine learning (ML) and deep learning (DL). TPUs focus on optimizing tensor operations and matrix multiplication, while NPUs aim to emulate human neural networks in computation. Both accelerators can be implemented in FPGA or ASIC forms.

\textbf{FPGA} LLM accelerators implemented by FPGA form provide an intermediate ground between ASICs and GPUs, offering the flexibility of programmability and a certain level of optimization for LLM computations simultaneously. They can adapt to new LLM architectures or algorithms through relatively flexible hardware reconfiguration and have relatively low power consumption.

\textbf{ASIC} ASIC-based LLM accelerators are custom-designed chips tailored to LLM computations. Unlike FPGAs, ASICs are optimized for specific tasks, offering excellent efficiency. They enable optimized memory usage in LLM computations and generally have lower power consumption than GPUs.

\begin{table*}[]
    \centering
    \caption{Comparison of LLM Accelerator Hardware Specifications}
    \resizebox{.95\textwidth}{!}{
        \begin{threeparttable}           
            \label{tab:hardware_spec}
            \begin{tabular}{@{}ccccccccc@{}}
                \toprule
                \multirow{2.4}{*}{\textbf{Accelerator}} & 
                \multirow{2.5}{*}{\textbf{\begin{tabular}[c]{@{}c@{}}Technology \\ Node\end{tabular}}} & 
                \multirow{2.5}{*}{\textbf{\begin{tabular}[c]{@{}c@{}}TDP\\ (W)\end{tabular}}} & 
                \multicolumn{3}{c}{\textbf{Memory}} & 
                \multirow{2.5}{*}{\textbf{\begin{tabular}[c]{@{}c@{}}Release \\ Date\end{tabular}}} &
                \multirow{2.4}{*}{\textbf{Latest Release}} & 
                \multirow{2.4}{*}{\textbf{Supported Operations}} \\ 
                \cmidrule(lr){4-6} 
                 &  &  & \textbf{Type} & \textbf{Capacity} & \textbf{Bandwidth} &  &  &  \\ 
                \midrule
                NVIDIA H100             & 4nm          & 700   & HBM3          & 80GB          & 3.35TB/s          & 2022 & GB200 & Training, Inference \\
                AMD MI300X              & 6nm/5nm    & 750   & HBM3          & 192GB         & 5.2TB/s           & 2023 & - & Training, Inference \\
                Cerebras WSE-2          & 7nm          & 20k   & SRAM          & 40GB          & 20PB/s            & 2021 & WSE-3 & Training, Inference \\
                Google TPU v4           & 7nm          & 192   & HBM2          & 32GB          & 1.2TB/s           & 2020 & TPU v6 (Trillium) & Training, Inference \\
                Graphcore IPU           & 7nm          & 300   & SRAM          & 0.9GB         & 65TB/s            & 2020 & - & Training, Inference \\
                Groq LPU                & 14nm              & 300   & SRAM          & 230MB         & 80TB/s            & 2024 & - & Inference \\
                Intel Gaudi 3           & 5nm          & 900   & HBM2E         & 128GB         & 3.7TB/s           & 2024 & - & Training, Inference \\
                SambaNova SN40L         & 5nm          & -     & HBM3, DDR5    & 64GB, 1.5TB   & 2TB/s, 200GB/s    & 2023 & - & Training, Inference \\
                Tenstorrent Grayskull   & 12nm              & 200   & LPDDR4        & 8GB           & 118.4GB/s         & 2024 & Wormhole & Inference \\ 
                \bottomrule
            \end{tabular}
        \end{threeparttable}
    }
\end{table*}

%-------------------------------------------------------------------------------
\subsection{Requirements for Accelerators}
%-------------------------------------------------------------------------------

LLM accelerators must meet various requirements beyond efficiently processing Transformer blocks, such as multi-head attention and feedforward operations, widely used in LLM training and inference.

\textbf{Low Power Consumption.} LLM accelerators need high throughput with lower power consumption than GPUs, requiring hardware-software co-design for energy-efficient computations. Optimization of memory and host-accelerator interfaces, along with compiler technologies, is essential.

\textbf{Low Latency.} Reducing inference latency is vital for LLM services. Techniques like tiling and pipelining, which break down LLM computations into smaller units for optimal processing, are under study~\cite{xia2023flash}. Methods to optimize data transfer between accelerators and memory, or between accelerators, are also proposed~\cite{moon2024lpu}.

\textbf{High Memory Capacity.} Models like Llama 3.1~\cite{dubey2024llama}, with up to 405 billion parameters, demand significant memory. Inference requires 1,944 GB or 972 GB for 32-bit or 16-bit precision, equating to about 24 or 12 NVIDIA H100 80GB GPUs. Efficient memory management also requires optimization in storing and loading the model's Queries, Keys, and Values~\cite{luohe2024keep, dao2022flashattention}.

\textbf{Support for Mixed Precision.} To reduce the memory required during LLM training and inference, methods like quantization~\cite{hooper2024kvquant, zhao2024atom} are applied to reduce memory usage. While 4-bit and 8-bit formats are used, mixed precision is needed for operations like addition in Transformer blocks.

\textbf{Support for Parallel Processing.} As LLMs become larger, executing models on a single accelerator becomes more challenging. Parallel and distributed training and inference techniques using multiple accelerators are necessary~\cite{brakel2024model}.

%-------------------------------------------------------------------------------
\section{Review of LLM Accelerators}
\label{sec:3}
%-------------------------------------------------------------------------------

For the analysis of LLM accelerators, we selected NVIDIA H100~\cite{hopper2023nvidia}, AMD MI300X~\cite{smith2024interconnect, smith2024amd}, Cerebras WSE-2~\cite{lie2023cerebras}, Google TPU v4~\cite{jouppi2023tpu, zu2024resiliency}, Graphcore IPU~\cite{perez2023training, balancca2024scalify}, Groq LPUP~\cite{abts2022software, abts2020think}, Intel Gaudi 3~\cite{amaro2023voyager, gaudi2024wp}, SambaNova SN40L~\cite{prabhakar2024sambanova}, and Tenstorrent Grayskull~\cite{thuning2024attention}. 
% We analyzed proprietary LLM accelerators that have been published in research papers or white papers. This section discusses the hardware and software characteristics of the selected LLM accelerators.

%-------------------------------------------------------------------------------
\subsection{Hardware Characteristics}
%-------------------------------------------------------------------------------

%-------------------------------------------------------------------------------
% \subsubsection{Hardware Specifications}
\subsubsection{Memory and Performance}
%-------------------------------------------------------------------------------

Table~\ref{tab:hardware_spec} shows specifications such as memory and power consumption for each LLM accelerator. TDP is the Thermal Design Power, GB denotes Gigabytes, and TB denotes Terabytes. Latest Release refers to the most recently launched hardware.

\textbf{NVIDIA H100 (SXM)} is widely used for LLM training and inference, it consists of 132 Streaming Multiprocessors (SMs), each comprising Tensor Cores, caches, etc.

\textbf{AMD MI300X} uses a chiplet architecture~\cite{li2020chiplet}, with eight Accelerator Complex Dies (XCDs) for acceleration.

\textbf{Cerebras WSE-2} is a wafer-scale chip with an area over 55 times larger than a typical GPU chip (WSE-2: 46,255 mm$^2$; NVIDIA A100: 826 mm$^2$), containing 850,000 cores. It provides 40 GB of memory using SRAM but has disadvantages like extremely high power consumption exceeding 20 kW.

\textbf{Google TPU v4} consists of two TensorCores per chip. TPUs follow an optically reconfigurable network architecture for large-scale machine learning tasks, enabling supercomputer-scale distributed processing by connecting up to 4,096 chips.

\textbf{Graphcore IPU (Colossus MK2 IPU Processor)} consists of 1,472 independent processor cores specialized for large-scale parallel processing. Each IPU is equipped with 900 MB of SRAM, with SRAM allocated to each processing unit called a tile within the IPU, providing fast data communication based on an In-Processor Memory architecture.

\textbf{Groq LPU} composed of Software-Defined Tensor Streaming Processor (TSP) for large-scale computation. It can scale up to 10,440 TSPs through network interconnects. Equipped with 230 MB of SRAM, hundreds of LPUs are required to run models like Llama 70B, leading to significant power consumption and operational cost issues.

\textbf{Intel Gaudi 3} offers high parallel processing performance through eight Matrix Multiplication Engines (MMEs) and 64 Tensor Processor Cores (TPCs). It is equipped with 128 GB of HBM2E, supporting over 3 TB/s of bandwidth.

\textbf{SambaNova SN40L} consists of 1,040 Distributed Pattern Compute Units (PCUs) for computation. Unlike other LLM accelerators, it features a three-tier memory system comprising HBM, DRAM, and SRAM. This memory configuration supports performance improvement through high parallelism by fusing hundreds of complex operations into a single kernel call using streaming dataflow parallelization.

\textbf{Tenstorrent Grayskull} utilizes a RISC-V-based architecture and is designed to intelligently process and move data through up to 120 compute cores called Tensix Cores via an on-chip Network-on-Chip (NoC) structure.

The performance of each LLM accelerator is shown in Table~\ref{tab:flops}. TFLOPS indicates Tera Floating Point Operations per Second; PFLOPS indicates Peta Floating Point Operations per Second; TOPS indicates Tera Operations per Second. Missing values are denoted by a dash (-). Each value represents the performance of tensor or matrix operations and may vary depending on the hardware development version and documentation.

\begin{table}[]
    \centering
    \caption{Comparison of LLM Accelerator Performance}
    \resizebox{.49\textwidth}{!}{
        \begin{threeparttable}            
            \label{tab:flops}
            \begin{tabular}{@{}ccccccc@{}}
                \toprule
                \multirow{2.4}{*}{\textbf{Accelerator}} & \multicolumn{6}{c}{\textbf{Data Type}} \\ \cmidrule(l){2-7} 
                 & \textbf{FP8} & \textbf{FP16} & \textbf{BF16} & \textbf{INT8} & \textbf{FP32} & \textbf{FP64} \\ 
                \midrule
                NVIDIA H100          & 3,958 TFLOPS  & 1,979 TFLOPS & 1,979 TFLOPS & 3,958 TOPS & 67 TFLOPS & 34 TFLOPS \\
                AMD MI300X           & 20.9 PFLOPS   & 10.5PFLOPS  & 10.5 PFLOPS  & 20.9 POPS  & 1.3 PFLOPS & 1.3 PFLOPS \\
                Cerebras WSE-2       & -             & 75 PFLOPS   & -            & -          & -          & - \\
                Google TPU v4        & -             & -           & 275 TFLOPS   & 275 TOPS & -          & - \\
                Graphcore IPU        & -             & 250 TFLOPS  & -            & -          & 62 TFLOPS  & - \\
                Groq LPU             & -             & 205 TFLOPS  & -            & 820 TOPS   & -          & - \\
                Intel Gaudi 3        & 1,835 TFLOPS  & 459 TFLOPS  & 1,835 TFLOPS & -          & 229 TFLOPS & - \\
                SambaNova SN40L      & -             & -           & 638 TFLOPS   & -          & -          & - \\
                Tenstorrent Grayskull& 332 TFLOPS    & 92 TFLOPS   & -            & -          & -          & - \\ 
                \bottomrule
            \end{tabular}
            % \begin{tablenotes}
            %     \footnotesize
            %     \item[*] Note: TFLOPS indicates Tera Floating Point Operations per Second; PFLOPS indicates Peta Floating Point Operations per Second; TOPS indicates Tera Operations per Second. Missing values are denoted by a dash (-).
            %     \item[**] Note: Each value represents the performance of tensor or matrix operations and may vary depending on the hardware development version and documentation.
            % \end{tablenotes}
        \end{threeparttable}
    }
\end{table}

\begin{table*}[]
    \centering
    \caption{Comparison of LLM Accelerator Interconnection}
    \resizebox{\textwidth}{!}{
        \begin{threeparttable}   
            \label{tab:interconnect}
            \begin{tabular}{@{}ccccccc@{}}
                \toprule
                \multirow{2.4}{*}{\textbf{Accelerator}} & \multicolumn{2}{c}{\textbf{Chip-to-chip}} & \multicolumn{2}{c}{\textbf{Card-to-card (Node)}} & \multicolumn{2}{c}{\textbf{Node-to-node}} \\ 
                \cmidrule(l){2-7} 
                 & \textbf{Type} & \textbf{Bandwidth} & \textbf{Type} & \textbf{Bandwidth} & \textbf{Type} & \textbf{Bandwidth} \\ 
                \midrule
                NVIDIA H100     & NVLink & 900 GB/s & NVLink/NVSwitch & 900 GB/s & InfiniBand & 400 Gb/s \\
                AMD MI300X      & Interposer bridge & - & Infinity Fabric & 128 GB/s & Infinity Fabric/PCIe Gen 5 & - \\
                Cerebras WSE-2  & SwarmX Fabric & 220 Pb/s & Custom Interconnect & - & - & - \\
                Google TPU v4   & Inter-Core Interconnect Link & - & Optical Circuit Switches (OCSes) & 50 GB/s & OCSes & 50 GB/s \\
                Graphcore IPU   & IPU-Links & 320 GB/s & GW-Link & 100 Gb/s & Sync-Link & 100 Gb/s \\
                Groq LPU        & Chip-to-chip (C2C) links  & 30 Gb/s & C2C links (Ethernet/QSFP) & - & C2C links (Ethernet/QSFP) & - \\
                Intel Gaudi 3   & Interposer bridge & - & RoCE (PAM 4 SerDes) & 112 Gb/s & RoCE (PAM 8 SerDes) & 800 Gb/s \\
                SambaNova SN40L & Die-to-die direct connect & - & Peer-to-Peer (P2P) Interface & - & Top Level Network (TLN) & - \\
                Tenstorrent Grayskull& NoC (Torus) & 192 GB/s & PCIe & - & - & - \\ 
                \bottomrule
            \end{tabular}
            % \begin{tablenotes}
            %     \footnotesize
            %     \item[*] Note: Bandwidth is specified in GB/s or Gb/s. Missing values are denoted by a dash (-).
            %     \item[**] Note: Bandwidth may vary according to the hardware version and documentation of each manufacturer. The values for Nodes might represent total bandwidth, as individual bandwidth distinctions are not always specified by the manufacturers.
            % \end{tablenotes}
        \end{threeparttable}
    }
\end{table*}

%-------------------------------------------------------------------------------
\subsubsection{Interconnections}
%-------------------------------------------------------------------------------

Another crucial aspect of LLM accelerators is interconnection. As the size of LLMs grows, it becomes challenging to process them with a single accelerator, making interconnection technologies within and between accelerators extremely important. The interconnection of each LLM accelerator are shown in Table~\ref{tab:interconnect}. Bandwidth is specified in GB/s or Gb/s. Missing values are denoted by a dash (-). Chip-to-chip refers to interfaces or implementations between chips within a single accelerator or package, Card-to-card represents interconnections between accelerators, and Node refers to a system with multiple accelerators. Node-to-node represents interconnections among nodes. Bandwidth may vary according to the hardware version and documentation of each manufacturer. The values for Nodes might represent total bandwidth, as individual bandwidth distinctions are not always specified by the manufacturers.

% Most LLM accelerators provide various interconnections for scaling up and scaling out. With the advancement of chiplet architectures, chip-to-chip communication is sometimes supported using interposer bridges. Each LLM accelerator uses Ethernet or private interfaces instead of PCIe for high-speed communication.

\textbf{NVIDIA H100 (SXM)} supports interconnection using proprietary NVLink, NVSwitch, and InfiniBand (Mellanox/NVIDIA). NVLink is used for communication between GPUs, generally offering higher bandwidth than PCIe. NVSwitch is a switch that interconnects large GPU clusters, managing and processing large-scale network traffic. InfiniBand is a high-speed communication technology between computing nodes, supporting Remote Direct Memory Access (RDMA), and is used in High-Performance Computing (HPC) and data centers. NVIDIA provides the NVIDIA Collective Communications Library (NCCL) to support optimal parallel processing for interconnecting multiple GPUs.

\textbf{AMD MI300X} uses the proprietary Infinity Fabric to support high-bandwidth interconnection. Infinity Fabric Mesh provides accelerators with bidirectional links of 128 GB/s. AMD also offers the ROCm Communication Collectives Library (RCCL) for communication between multiple LLM accelerators.

\textbf{Google TPU v4} supports low-latency data transfer between chips through the Inter-Chip Interconnect (ICI). Multiple connected TPUs are called a TPU Pod, and an Optical Circuit Switch (OCS) is used to reconfigure large-scale network topologies for communication between TPU Pods. OCS offers lower latency and higher data transfer capacity than traditional electronic switches, enhancing the scalability and flexibility of TPU clusters.

\textbf{Intel Gaudi 3} supports high-speed data interconnection through integrated Ethernet-based RDMA (RDMA over Converged Ethernet, RoCE) network technology. RoCE implements the RDMA protocol over Ethernet, enabling low-latency data transfer through direct data transmission between accelerators. Gaudi 3 offers high scalability and communication efficiency through large-scale parallel data transmission via RDMA NICs. Intel also provides optimization technologies for interconnecting multiple devices through the Intel oneAPI Collective Communications Library (oneCCL).

\textbf{Tenstorrent Grayskull} allows data exchange between cores through a 2D bi-directional torus network. Designed based on a dataflow architecture, it offers high efficiency in parallel computing environments. However, PCIe is used for card-to-card communication.

Other accelerators like Graphcore IPU support accelerator clustering through IPU-Link and GW-Link, and Cerebras WSE-2, Groq LPU, and SambaNova SN40L support communication between multiple accelerators using custom interconnections.

\begin{table}[]
    \centering
    \caption{Comparison of LLM Accelerator Software Support}
    \resizebox{0.49\textwidth}{!}{
        \begin{threeparttable}
            \label{tab:software}
            \begin{tabular}{@{}c>{\centering\arraybackslash}p{3cm}ccccccc@{}}
                \toprule
                \multirow{2.4}{*}{\textbf{Accelerator}} & \multicolumn{2}{c}{\textbf{Dedicated Software}}  & \multicolumn{5}{c}{\textbf{Common Framework}} \\ \cmidrule(l){2-8} 
                                                      & \textbf{Framework}        & \textbf{SDK/Library} & \textbf{Tensorflow} & \textbf{PyTorch} & \textbf{JAX} & \textbf{vLLM} & \textbf{DeepSpeed} \\ \midrule
                NVIDIA H100                           & TensorRT-LLM              & NVIDIA CUDA          & \ding{51}  & \ding{51}   & \ding{51}  & \ding{51}  & \ding{51}  \\
                AMD MI300X                            & MosaicML, LUMI, Lamini    & AMD ROCm             & \ding{115} & \ding{115}  & \ding{53}  & \ding{51}  & (MI200)    \\
                Cerebras WSE-2                        & Cerebras Inference        & CSL                  & \ding{115} &  \ding{115} & \ding{53}  & \ding{53}  & \ding{115} \\
                Google TPU v4                         & JAX                       & XLA                  & \ding{51}  & \ding{115}  & \ding{51}  & \ding{51}  & \ding{53}  \\
                Graphcore IPU                         & Poplar                    & Poplar SDK           & \ding{115} & \ding{115}  & \ding{115} & \ding{115} & \ding{53}  \\
                Groq LPU                              & GroqWare/GroqFlow         & LPU Inferene Engine  & \ding{115} & \ding{115}  & \ding{53}  & \ding{53}  & \ding{53}  \\
                Intel Gaudi 3                         & -                         & HCCL                 & \ding{115} & \ding{115}  & \ding{53}  & \ding{115} & (Gaudi2)   \\
                SambaNova SN40L                       & SambaNova Suite/SambaFlow & SambaNova SDK        & \ding{115} & \ding{115}  & \ding{53}  & \ding{53}  & \ding{53}  \\
                Tenstorrent Grayskull                 & TT-Buda                   & TT-Metalium          & \ding{115} & \ding{115}  & \ding{53}  & \ding{53}  & \ding{53}  \\ 
                \bottomrule
            \end{tabular}
            % \begin{tablenotes}
            %     \footnotesize
            %     \item[*] Note: O indicates official support within the framework, while (O) signifies unofficial or experimental support. When a product name is enclosed in parentheses, it indicates that the hardware is supported.
            % \end{tablenotes}
        \end{threeparttable}
    }
\end{table}

%-------------------------------------------------------------------------------
\subsection{Software Characteristics}
%-------------------------------------------------------------------------------

An important aspect of LLM accelerators is compatibility with existing LLM software. Frameworks like TensorFlow, PyTorch, and JAX, commonly used in machine learning and deep learning, as well as specialized frameworks for LLM development like vLLM~\cite{kwon2023efficient} and DeepSpeed~\cite{rasley2020deepspeed}, are being used. Table~\ref{tab:software} shows the software support status of proprietary LLM accelerators. \ding{51} indicates official support within the framework, while \ding{115} signifies unofficial or experimental support. When a product name is enclosed in parentheses, it indicates that the hardware is supported.

NVIDIA H100, one of the representative LLM accelerators, is officially supported by various frameworks due to its widespread use in LLMs. NVIDIA provides dedicated frameworks like TensorRT-LLM and performs computation acceleration through the NVIDIA CUDA library.

LLM accelerators like AMD MI300X, Cerebras WSE-2, and Google TPU v4 also support various proprietary frameworks and libraries for hardware acceleration. LLM accelerators not officially supported by existing LLM frameworks distribute modified versions of existing frameworks' source codes, adding compilers and model conversion codes to support their hardware. They may also use methods that convert codes or models developed in existing frameworks into compatible models like ONNX~\cite{jin2020compiling} and connect them to their proprietary frameworks.

To support various LLMs, each LLM accelerator optimizes foundation models to fit their hardware and distributes them. However, since it's challenging to support all models, differences arise in the models that can be trainining and inference by each accelerator.

%-------------------------------------------------------------------------------
\section{Challenges and Future Directions for LLM Accelerators}
\label{sec:4}
%-------------------------------------------------------------------------------
In this section, we discuss the challenges and considerations based on the proprietary LLM accelerators explained above.

\textbf{Memory} As the scale of LLMs increases, the required memory capacity also grows. While HBM has been adopted to meet these needs, the complex manufacturing process of HBM leads to increased prices for LLM accelerators. Due to the continuously increasing size of LLMs, new solutions are needed. Although LLM accelerators equipped with SRAM or LPDDR have been proposed as alternatives to HBM, these may not suffice for future demands.

% To efficiently perform LLM computations, architectures like Processing-in-Memory (PIM) are being explored in hardware based on traditional NPUs or accelerator units. PIM combines compute units with memory to reduce latency related to memory storage and retrieval. Currently, DRAM-based UPMEM~\cite{gomez2021benchmarking} and GDDR6-based SK Hynix AiMX~\cite{kwon2022system, lee2024cost} are being developed. However, issues remain regarding JEDEC standard compliance, low memory capacity, and usability.

\textbf{Low Power Consumption} The LLM accelerators analyzed in this paper showed TDPs ranging from 192W to 20kW. In particular, AMD MI300X and Intel Gaudi 3 exhibited power consumption of 50–200W higher than the NVIDIA H100. Additionally, the total system power consumption increases sharply as multiple accelerators are required for LLM services.

% According to a study analyzing GPU power consumption~\cite{son2023thermal}, in a package consisting of six HBMs and one GPU, the HBM and GPU accounted for approximately 6\% and 91\% of total power consumption, respectively. While having large memory through HBM is important, optimizing the power consumption of compute units like GPUs that perform LLM computations is necessary. Innovative cooling solutions and energy-efficient hardware design are potential areas for future research.

\textbf{Scalability }Most LLM accelerators use interconnections developed by each manufacturer. Executing massive LLM models on a single accelerator is challenging; therefore, interconnection technologies are crucial for scaling up by connecting multiple accelerators and scaling out by expanding these connected accelerator groups. 

% For instance, the Llama 3 405B model was trained using 16,000 NVIDIA H100s~\cite{dubey2024llama}. While hardware interconnection is important, supporting efficient parallel processing of LLMs at the software level is equally critical. Developing standardized interconnects and improving distributed computing frameworks could enhance scalability. 

\textbf{Software Compatibilit}y Most LLM accelerators use proprietary software stacks. Providing a seamless development environment is essential as many users rely on frameworks commonly used with GPUs. Furthermore, while support from existing frameworks for LLM accelerators is important, the most critical aspect is compiler technology that allows developed models to perform optimally on LLM accelerator hardware. 

% Technologies like MLIR~\cite{lattner2021mlir} and Triton~\cite{banerjee2023triton} are being researched to serve as common solutions across various LLM accelerators. MLIR is a compiler infrastructure that provides a unified intermediate representation for different hardware targets, facilitating optimization and code generation. Triton is a language and compiler for writing efficient GPU code with minimal effort, aiming to simplify the development of high-performance kernels. 

%-------------------------------------------------------------------------------
\section{Conclusions}
\label{sec:5}
%-------------------------------------------------------------------------------

In this paper, we analyzed the hardware and software characteristics of commercial accelerators for LLMs and discussed the challenges they face. By comparing the specifications and performance of major accelerators, we identified current technological trends and issues such as memory capacity, power consumption, scalability, and software compatibility. Future research requires innovative hardware designs and software optimization technologies to address these challenges, focusing on enhancing energy efficiency and scalability while ensuring software compatibility.

%-------------------------------------------------------------------------------
\section*{Acknowledgments}
%-------------------------------------------------------------------------------

This work was supported by Institute of Information \& communications Technology Planning \& 
Evaluation (IITP) grant funded by the Korea government (MSIT) (No. RS-2024-00402898, 
Simulation-based High-speed/High-Accuracy Data Center Workload/System Analysis Platform)

%-------------------------------------------------------------------------------
% \bibliographystyle{plain}
\bibliographystyle{unsrt}
\bibliography{aicomps}

%%%%%%%%%%%%%%%%%%%%%%%%%%%%%%%%%%%%%%%%%%%%%%%%%%%%%%%%%%%%%%%%%%%%%%%%%%%%%%%%
\end{document}